\documentclass[12pt]{article}
\pdfoutput=1

\usepackage{amsmath,amssymb,amscd}
\usepackage{listings}
\usepackage{caption}
\usepackage{dsfont}
\usepackage{slashed}
\usepackage{color}
\usepackage{ulem}
\usepackage[pdftex]{graphicx}
\usepackage{epstopdf}
\usepackage{subfigure}
\usepackage{epsfig}
\usepackage{listings}
\usepackage{caption}
\usepackage{cite}
\usepackage{multirow}
\usepackage{upgreek}

\newcommand{\beq}{\begin{equation}}
\newcommand{\eeq}{\end{equation}}
\newcommand{\nbea}{\begin{align*}}
\newcommand{\neea}{\end{align*}}
\newcommand{\nbeq}{\begin{equation*}}
\newcommand{\neeq}{\end{equation*}}

\newcommand{\up}{\mathbf{u_p}}
\newcommand{\ug}{\mathbf{u}}
\newcommand{\re}{{\rm Re}}
\newcommand{\dsp}{d_{\rm dsp}}

\usepackage{multirow}
\usepackage{array}
\newcolumntype{M}[1]{>{\centering\arraybackslash}m{#1}}
\newcolumntype{N}{@{}m{0pt}@{}}

\begin{document}

\baselineskip=21pt

\begin{center}

{\large {\bf Study of Air Curtain in Context of Individual Protection from Exposure to Coronavirus (SARS-CoV-2)
Contained in Cough-Generated Fluid Particles}}

\vskip 0.5cm

{\bf Alexander S. Sakharov}\textsuperscript{a,b}~\footnote{Corresponding author (Alexandre.Sakharov@cern.ch)} and
{\bf Konstantin Zhukov}\textsuperscript{b,c}~\\

\vskip 0.2in

{\small {\it

\vspace{0.25cm}
\textsuperscript{a}Physics Department, Manhattan College\\
{\mbox 4513 Manhattan College Parkway, Riverdale, NY 10471, United States of America}\\
\vspace{0.25cm}
\textsuperscript{b}Experimental Physics Department, CERN, CH-1211 Gen\`eve 23, Switzerland}\\
\vspace{0.25cm}
\textsuperscript{c}P.N. Lebedev Physical Institute of the Russian Academy of Sciences, 53 Leninskiy Prospekt, 119991, Moscow, Russia \\}

\vskip 0.2in

{\bf Abstract}

\end{center}

\baselineskip=18pt \noindent


The~ongoing respiratory COVID-19 pandemic has heavily impacted the social
and private lives of the majority of the global population.
This infection is primarily transmitted via virus-laden fluid particles (i.e., droplets and aerosols)
that are formed in the respiratory tract of infected individuals and expelled
from the mouth in the course of breathing, talking, coughing, and sneezing.
To mitigate the risk of virus transmission, in many places of the world, the public has been
asked or even obliged to use face covers. It is plausible that in the years ahead we will see the use of
face masks, face shields and respirators become a normal practice in our life.
However, wearing face covers is uncomfortable in some situations, like, for example,
in summer heat, while staying on beaches or at hotel swimming pools, doing exercises in gyms, etc.
Also, most types of face cover become contaminated with time and need to be
periodically replaced or disinfected. These nuisances are caused by the fact that
face covers are based on material barriers, which prevent inward and outward
propagation of aerosol and droplets containing the pathogen.
Here, we study a non-material based protection barrier created by
a flow of well directed down stream of air across the front of the open face.
The~protection is driven by dragging virus-laden particles inside the width of the air flow and hence, as
a consequence, displacing them away from their primary trajectories.
Applying well established gas-particle flow formalism,
we analyzed the dynamics of aerosols and droplets at different
regimes of the flow laying over the bodies of the fluid particles.
The~analysis allowed us to establish the rates of velocity gain of the fluid particles
of dimensions relevant for the pathogen transmissions, while they are crossing
the width of the air barrier.  On the basis of this analysis,
we provide a comprehensive study of the protection effectiveness of the
air barrier for a susceptible individual located indoor,
in an infected environment.
Our study shows that such, potentially portable, air curtains can effectively provide
both inward and outward protection and serve as an effective personal
protective equipment (PPE) mitigating human to human transmission of
virus infection like COVID-19.

\vskip 3mm

\vskip 5mm

\leftline{June 2020}

\vskip 10mm


\section{Introduction}
\label{intro}
The~Severe Acute Respiratory Syndrome Coronavirus-2 (SARS-CoV-2) has caused the COVID-19 pandemic disease.
The~virus of diameter 70--90~nm~\cite{virusSize_1} is transmitted from human to human,
primarily being carried by virus-laden fluid particles ejected
from the mouth of infected individuals. In general, COVID-19 infection has multiple mode of transmission.
It may spread by respiratory droplets
and aerosols~\cite{surfaceSurvCOVID_1,modAir,aerosolTrans_1}, via direct contact such as hand shake or trough indirect contact via
contaminated surfaces~\cite{surfaceSurvCOVID_1}.  The~infectious dose for SARS-CoV-2 is not known.
Extrapolating from studies of other viruses where
more data are available, one might expect the threshold to vary from a few
tens to a few thousand virus exemplars~\cite{infDose1}. From these data, it is also likely that the larger the
dose of virus an individual is
exposed to, the higher the likelihood of infection occurring. Whether the disease severity
correlates with the size of the infectious dose is also unclear. However, based on animal experiments performed with a variety of
viruses~\cite{lethal}, it is reasonable to assume that this is the case. SARS-CoV-2 can remain infective in
aerosols for 3~h and on surfaces for 72~h in laboratory conditions~\cite{surfaceSurvCOVID_1}.
However, evidence about acquiring the infection via contaminated surfaces is scarce~\cite{surfaceSurvCOVID_2}.

Fluid particles responsible for the respiratory transmission are represented by aerosol and
droplets~\cite{surfaceSurvCOVID_1,modAir,aerosolTrans_1}. In this context, aerosol particles
may contain exemplars of virus pathogen, epithelial and other cells or remnants of those,
natural electrolytes and other substances from mucus and saliva, and water which typically evaporates
quite fast, depending on the relative humidity of the surrounding air.
Staying in the air for a long enough time (minutes or hours), aerosol particles
of dimensions $\lesssim$5~{$\upmu$m} can be
inhaled and airborne transferred over long distance in an indoor environment~\cite{dropletDry_1,dopletDry}. Large droplets
$\gtrsim$200~${\upmu}$m stay airborne only for few seconds. Their movement
mainly follows ballistic trajectories and to less extend the ambient air
flow. Such droplets settle to the ground or other surrounding surface. Smaller droplets,
say $\lesssim$5~{$\upmu$m} would evaporate in less than 3 s, at typical
indoor relative humidity ${\rm RH=50\%}$.
Since the sedimentation time of such a droplet
is about 30~s, it would evaporate completely before reaching the background.
Actually, the evaporation time for
water droplets smaller than  $80$~{$\upmu$m}  does not
exceed their sedimentation time~\cite{modAir}.
The~drying process would leave residuals of droplets after drying to moisture in equilibrium
with ambient air. These residuals, so called droplet nuclei, are also airborne
transferred aerosols. Aerosols particles ($\lesssim$10~{$\upmu$m}) can float on the air and
spread a significant distance ($\mathcal{O} (10)$~m~\cite{modAir}) following air flow streams, in
particular after being dried~\cite{coughSim_1,modAir}.

The~number density, velocity, and size distributions of fluid particles ejected in the course of expiratory
events have important implications for transmission~\cite{sneezDrop_1,coughSim_1,modAir,speech_1}.
A single sneeze action can generate $\mathcal{O} (10^4)$ fluid particles moving with velocities up to $20$~m/s~\cite{sneezDrop_1}.
Coughing generates 10--100 fewer droplets than sneezing at lower velocity  $\lesssim 10$~m/s~\cite{coughSim_1}.
Along with coughing and sneezing, speaking also
plays a significant role in spreading the virus contained particles~\cite{speech_1}.
In particular, as it is normally done continuously over a longer period of time,
the contribution into virus transmission can be significant.
Moreover, the rate of particle production in the course of normal human speech
is in positive correlation with the loudness (amplitude) of vocalization, ranging from
approximately 1 to 50 particles per second~\cite{speech_1}. Needless to say, that
SARS-CoV-2 infected individuals can spread the disease before the onset of clinical symptoms and
the infection will be mostly transmitted in course of speaking generated virus-laden particles.
Measured ejected particle size spans four orders of magnitude from
$\simeq$0.1 to $\simeq$1000~${\upmu}$m~\cite{speech_1,sizeDistr1}.

Face cover masks are currently treated as the best accepted personal protective equipment
(PPE) in mitigating aerosol dispersal. In conditions of gradual softening of the
pandemic outbreak measures, in many places, public has been asked or even
obliged to wear masks. Thus, face masks, in some sense, become a norm in our lives.
However, wearing face masks may become uncomfortable in some situations, like, for example,
in summer heat, while staying on beaches or at hotel swimming pools, doing exercises in gyms, etc.
Also, if an individual has a chronic respiratory condition such as asthma or
chronic obstructive pulmonary disease (COPD), covering
her/his mouth  and nose can be especially challenging. The~material barrier of the mask makes it
harder to take in air. It also traps some carbon dioxide as it is exhaled,
which means that one ends up breathing in air that is warmer and moister.
Unfortunately, that sensation of having trouble breathing in a mask might
get even worse in summer time. Many people with chronic lung conditions find
it harder to breathe in hot, humid air
(though some others fare worse when the weather is cold and dry).

The~focus of the paper is understanding the physics that underpins
the effectiveness of a possible non-material personal defense against airborne
pathogens like SARS-CoV-2. The~barrier of such non-material protection
is supposed to be created by a flow of well directed, slightly inclined, down stream of air
which screens the front of the
open face in the manner of an air curtain.
The~protection mechanism is driven by the dragging
of virus-laden airborne particles inside the width of the flow and as
a consequence displacing them away from their primary trajectories.
Our study shows that such, potentially portable, air curtain can provide
both inward and outward protection and serve as an effective personal
protective equipment (PPE) mitigating from human to human transmission of
virus infection like COVID-19.

The~structure of the paper is as follows. In Section \ref{ppe}, we describe
the main elements of the setup of a potentially portable air curtain to
be used as PPE from airborne transmission of virus infection
like COVID-19. In Section \ref{dragRegimes}, we investigate how the
fluid particles gain their velocities being dragged by an air flow
in different regimes of the flow layer over bodies. In Section \ref{effectiveness},
we discuss the protection mechanism of the air barrier and validate its efficiency.
In Section \ref{concl}, we conclude and comment upon the results of our study.

\section{Air curtain as PPE to mitigate airborne virus \\ transmission}
\label{ppe}

\begin{figure}
\centering
\includegraphics[angle=0,scale=0.26]{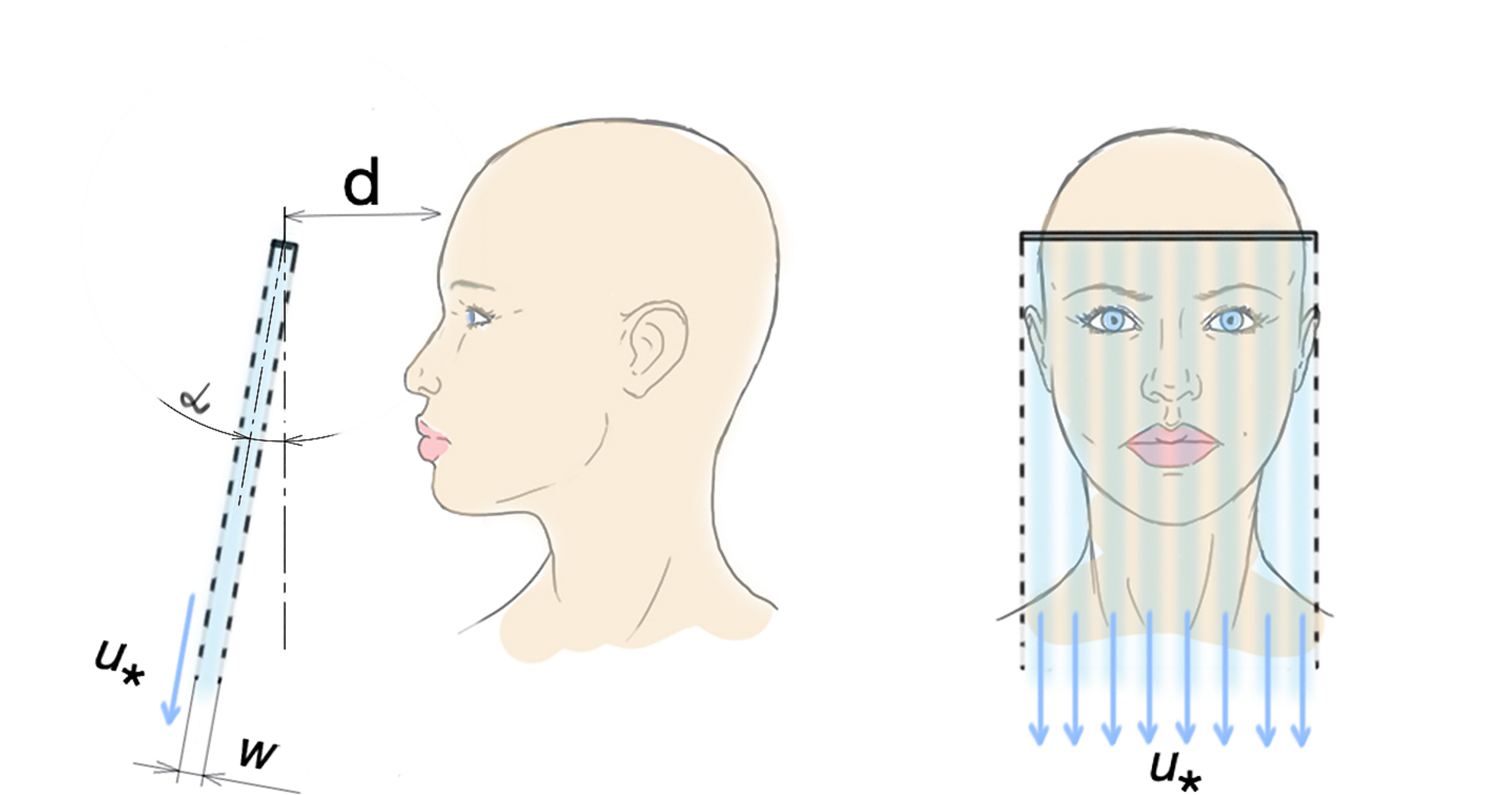}
\caption{Schematic representation of the barrier of air curtain formed by narrow
width, $w$, downward stream of velocity $u_*$, inclined at an angle $\alpha$
and localized at distance $d$ from the plane of the protected~face.}
\label{fig:face}
\end{figure}

We want to consider a portable air curtain  as a sort of dynamical PPE
from exposure of a recipient's mouth, nose and conjunctiva to virus-laden
fluid particles injected by infected individuals. The~air curtain is supposed
to be created as a flow of well directed downward stream of air across
the front of the open face, as schematically shown in~Figure \ref{fig:face}. This air stream
should provide a kind of dynamical barrier which steadily displaces
inward running virus-laden aerosolized particles that would otherwise
be inhaled by an uninfected person. In this setup, the outward protection
is also provided by stimulation of a ground settling of outgoing
virus-laden particles expelled by an infected person, so that sedimentation on surfaces
located substantially bellow the typical level of the face openings
would proceed faster than in usual conditions of indoors air background streams.
One~assumes that the curtain is formed
out of surrounding air and streamed downwards in a way that the flow
lines do not cross the plane of the face. For a sufficient downward
bending of the barrier penetrating inward running particles,
the plane of the air curtain should be inclined at a small angle $\alpha$
with respect to the vertical plane, as illustrated in Figure~\ref{fig:face}.

We assume that an appropriately designed system can create such a stream
across the entire height and width of the protected face. As a rough example of a particular
realization, one can imagine that the air flow is injected from the edge of a brim equipped with
a properly designed and instrumented air outlet which streams the air in a way that it forms a thin
air barrier of the horizontal shape defined by the geometry of the edge of the brim. A good coverage can be already
provided with dimensions and shape used for brims of the more-or-less standard baseball caps.
For~the instrumentation, we rely on current development of technology for construction of portable, low noisy
and low consuming pumping units, which will be able to create an air curtain
of about 1~cm thick with stable flow velocity $\lesssim$10~m/s
across an area of typical face dimensions.

In the current study, we investigate the effectiveness of a portable air curtain in
inward and outward protection  from transmission of respiratory
infections such as COVID-19. The~protection effectiveness,
in conditions of a typical parameters appropriate for creation of individual air curtain, mostly depends upon the ability of the air flow to band downward trajectories
of the fluid particles containing influenza, in a way to prevent them coming into contact with the
face surface of an uninfected individual.

In our setup, the degree of banding of the trajectories is defined
by the effectiveness of the air flow, streamed at certain velocity $u_*$ and angle $\alpha$,
to drag the fluid particles (aerosol and droplets) along the lines of the stream (see Figure~\ref{fig:face}),
while they are crossing the air curtain in transverse direction,
thus being immersed for some time
into the flow. The~drag force excreted to a particle
by moving media is defined as the component of the force
parallel to the direction of the motion of the flow
of the media~\footnote{For a review see~\cite{bookFluid_1,bookFluid_2}.}. 
This~force depends on the properties of the media, which is, in our case, the air at atmospheric
pressure, its velocity, the size of a particle immersed into the media and its material properties.
\mbox{At a given} velocity of the media, which we take $u_*=10$~m/s for our benchmark setup, the particles of
different sizes will gain their velocity in different regimes defined by the hydrodynamic conditions of the
flow layer over the body of the particles. These regimes are investigated in the following section.

\section{Dragging of virus-laden aerosols and droplets \\ by air flow}
\label{dragRegimes}

In case a particle velocity vector $\up$ is different from the air velocity vector, $\ug$, the drug force
exerted by the gas is given by Newtonian force
\beq
\label{newt1}
\mathbf{F_s}=-C_DS_m\frac{\rho (u_p -u)(\up -\ug)}{2},
\eeq
where $\rho$ is the density of the flow, $S_m$ is the maximal area of a body
in the plane perpendicular to the flow direction and $C_D$ is the drug (resistance) coefficient.
The~drug coefficient of a body depends on the velocity and the
viscosity of the drugging media (gas or liquid) and is defined by the
Reynolds~number
\beq
\label{Re1}
\re =\frac{\rho |u_p-u|D}{\mu},
\eeq
where $D$ and $\mu$ are the dimension of the body and the dynamical viscosity of
media. The~Reynolds number totally defines the mode of the flow layer over the body
and thus the law of resistance. Depending on the value of $Re$ one defines the three
regimes' flow layers over the a body, namely, the laminar regime ($0<{\rm Re}<1$), the intermediate regime, i.e., a transition from laminar to
turbulent  (\mbox{$1<{\rm Re}<700$}), and the turbulent regime  (${\rm Re}>700$)~\cite{bookFluid_1,bookFluid_2}.
The~equation describing the resistance (drugging) response of a media to transport of a particle
depends on the regime of the flow laying over the particle, thus it depends
on the Reynolds number.

As discussed above, the size of particles ejected by a speaking, coughing or sneezing
person spans the range between $0.1$~${\upmu }$m and $\simeq$1000~${\upmu}$m depending on the circumstances
of the ejection.
Further, we assume that the particles are spherical with smallest size 100~nm~\cite{speech_1,sizeDistr1}.
In~our task, to understand the regimes of movement of particles with different sizes,
we estimate their Reynolds numbers in a steady state flow of air at atmospheric
pressure and velocity $u=10$~m/s, when a particle is initially at rest, $u_p=0$.
Throughout the paper, we use for the air density $\rho =1.205$~$\rm{kg/m^3}$ and
for its dynamical viscosity $\mu =1.81\times 10^{-5}$~$\rm{Pa\cdot s}$. For a particle
of size $D_L=1.5$~$\rm{\upmu}$m the Reynolds number is $\re =0.998$, which implies
that sub-micron particles should move in laminar regime.
Bigger particles, namely those of the size between $D_L$ and $D_I=1050$~${\upmu}$m,
for which   $\re =699$, should move in an intermediate regime. Eventually, particles with dimension $D_T$
exceeding $D_I$ will be drugged in a turbulent, so-called, auto-model regime.

Let us study in detail the drugging of the aerosol particles by a steady air flow in these three~regimes.

For the laminar regime, the drug (resistance) coefficient of a hard spherical body
obeys the Stocks~formula
\beq
\label{Cd1}
C_D=\frac{24}{\re}.
\eeq

Therefore, the resistance force for a Stocks particle reads
\beq
\label{stocks1}
F_S=3\pi\mu uD(u_p-u)
\eeq

In the range of Reynolds numbers between 1 and 700, for the dependence $C_D(\re )$ a number of empirical
approximations are used. For the turbulent regime $\re >700$ the drug coefficient is constant, in particular,
for a hard spherical body, $C_D=0.44$.

Let us consider the drugging of a disperse particle by a steady state one dimensional
flow of media of density $\rho $
propagating along the $x$ axis.
We model the particle by a hard spherical body of diameter $D$ made out of material of density $\rho_p$, so that
the mass of the particle is given by
\beq
\label{mp1}
m = \frac{\pi D^3}{6}\rho .
\eeq

Assume, that at $x=0$ the particle with initial velocity $u_{p0}$ gets injected into the flow and let us
evaluate the velocity of the particle with time $u_p(t)$.
The~equation of motion can be expressed as follows
\beq
\label{eqM1}
\frac{\pi D^3}{6}\rho_p\frac{d\up}{dt}=C_D\frac{\pi D^2}{4}\rho\frac{u-u_p}{2}(\ug -\up).
\eeq

In Stocks regime ($\re <1$), which implies
\beq
\label{Cd2}
C_D=\frac{24\mu}{\rho (u-u_p)D},
\eeq
equation (\ref{eqM1}) can be reduced to the form
\beq
\label{eqM2}
\frac{D^2\rho_p}{18\mu}\frac{d\up}{dt}=(\ug -\up).
\eeq
The~coefficient
\beq
\label{rel1}
t_{S*}=\frac{D^2\rho_p}{18\mu},
\eeq
in front of the derivative in Equation (\ref{eqM2}), is the dynamical relaxation time in Stocks regime, which
indicates the estimate of time scale needed for a drugged particle to reach the velocity of
the flow. Further, for the flow of constant velocity $u(x)=u_*$, one can reduce
the equation of motion to the dimensionless~form
\beq
\label{eqM3}
\frac{dy}{d\tau}=1-y,
\eeq
where $y=u_p/u_*$ and $\tau =t/t_*$.
Solving in Equation (\ref{eqM3}), we arrive to
\vspace{12pt}
\beq
\label{sol1}
y=1-(1-y_0)\exp (-\tau),
\eeq
where $y_0=u_{p0}/u_*$ is set for the initial condition.

In the intermediate regime of flow layer over a particle,
the drug (resistance) coefficient $C_D(\re)$ is given by the Klyachko
formula~\cite{bookFluid_2}
\beq
\label{Cd3}
C_{D({\rm I})}=\frac{24}{\re}+\frac{4}{\re^{1/3}},
\eeq
often used as a good empirical approximation.

Using $C_{D({\rm I})}$ instead of $C_D$ and introducing a new variable $z=\re^{1/3}$, one can re-write the equation
of motion (\ref{eqM1}) as follows
\beq
\label{eqM5}
\frac{dz}{d\tau_1}=-Hz(6+z^2),
\eeq
where $t_{1*}=\frac{\rho_pD^2}{\mu}$ is a time scale so that $\tau_1=t/t_{1*}$ and $H=\frac{\mu}{\rho_pD^2}$ is a distance scale. Integrating this equation at initial
conditions set to $\tau_1 =0$ and $z=z_0$ one arrives to
\beq
\label{sol3}
z=\sqrt{6}\left[\left(1+\frac{6}{z_0^2}\right)\exp (-12\tau_1)-1\right]^{-1/2}.
\eeq

Finally, after introduction of new variables $y=u_p/u$ and $\tau_{I} =-12\tau_1$, the solution takes the following form
\beq
\label{sol4}
y=1-6\sqrt{6}\frac{1-y_0}{\re_0}\left[\left(1+\frac{6}{\re_0^{2/3}}\right)\exp (\tau_I)-1\right]^{-3/2},
\eeq
where $\re_0=\frac{\rho uD}{\mu}(1-y_0)$. We notice, that the relaxation time is given by
\beq
\label{rel2}
t_{I*}=\frac{D^2\rho_p}{12\mu}.
\eeq

Finally, in the turbulent regime of flow layer over a particle, when $\re >700$, the drug coefficient is just a constant
$C_{D({\rm T})}=0.44$~\cite{bookFluid_1,bookFluid_2}. Thus, the equation of motion (\ref{eqM1}) takes the form
\beq
\label{eqM6}
\frac{4}{3}\cdot\frac{\rho D}{\rho_pC_{D({\rm T})}}\cdot\frac{d\up}{dt}=(\ug -\up)^2,
\eeq
where the coefficient $x_{T*}=\frac{4}{3}\frac{\rho D}{\rho_pC_{D({\rm T})}}$ has a dimension of length.
Introducing the characteristic time as
\beq
\label{rel3}
t_{T*}=\frac{x_{T*}}{u_*}=\frac{4}{3}\frac{\rho D}{u\rho_pC_{D({\rm T})}},
\eeq
one can rewrite Equation (\ref{eqM6}) in following form
\beq
\label{eqM7}
\frac{dy}{d\tau_T}=(1-y)^2,
\eeq
where we use $\tau_T=t/t_{T*}$. The~solution of Equation  (\ref{eqM7}), with
initial condition $y=y_0$, is given by following~expression
\beq
\label{sol5}
y=\frac{y_0+\tau_T|1-y_0|}{1+\tau_T|1-y_0|}.
\eeq

\section{Discussion}
\label{effectiveness}

Using the result of Section \ref{dragRegimes}, one can estimate the effectiveness of
the individual air flow barrier in providing inward and outward protection.

In our benchmark setup, described in Section \ref{ppe}, we use the barrier formed by
downward stream of air with width $w=1$~cm, which is well localized in
$d\approx 10$~cm~away from the surface of a protected face and provides a constant flow velocity $u_*=10$~m/s across vertical and horizontal
dimensions of the face, as shown in Figure~\ref{fig:face}. For the current theoretical study, to some extend, the distance $d$ is an arbitrary parameter. The~choice of its value can be regulated by other factors, not directly related to the effectiveness of the PPE under study. The~plane
of the barrier is inclined at an angle $\alpha$ with respect to the vertical plane (see Figure~\ref{fig:face}).
The~value of $\alpha$ will be discussed bellow.

A susceptible individual, while walking indoors, through an environment containing
virus-laden particles, has a high risk of exposure to the infection through the openings of
his face. The~particles themselves, being airborne-transmitted, experience motion
due to the environmental low speed  background air flows with maximal velocity $u_b=0.2$~m/s~\cite{modAir}.
Assuming the maximal indoor walking speed $u_w=1.5$~m/s, one infers that
an inward running particle, entering horizontally under the exterior plane of the air curtain,
will cross the width of the barrier within $t_w\gtrsim 6$~ms. Actually, $t_w$ is the time available
for the particle to gain velocity $u_p$ aligned with the direction of the barrier air flow, due
to its drag force. In course of drugging, within the inclined air barrier, the particle will gain
a horizontal outward velocity in amount of $u_{ph} = u_p\sin\alpha$. Thus, the inward (horizontally) running particle velocity
will be reduced by $u_{ph}$ or even inverted in its direction, depending on the value of the inclination angle $\alpha$.
If the gained vertical velocity is high enough to displace the particle down, namely sufficiently away from a
potential contact with mucus and conjunctiva of the host's face, the effectiveness of the protection is treated as
justified. The~fractional velocity gained by the particle, in different laying over regimes,
is given by Formulas (\ref{sol1}), (\ref{sol4}) and (\ref{sol5}).

As we have discussed in the introduction, the airborne virus transmission
is associated with droplets smaller than about $100$~${\upmu}$m that are suspended and
transported in an air current~\cite{dropletDry_1,dopletDry}.
These~droplets can evaporate in a few seconds~\cite{dropletDry_1,dopletDry}, so that droplet nuclei may be formed.
Sometimes, a vapour-rich, buoyant turbulent expiratory jet can slow down this evaporation~\cite{jet_1}.
The~droplet nuclei size spans range from sub-microns to about $10$~${\upmu}$m. Thus, as argued in
Section \ref{dragRegimes}, at a chosen velocity of the barrier flow, the gas-particles motion of this droplets size range is
driven by Stocks and intermediate regimes.

\begin{figure}
\centering
\includegraphics[scale=0.7]{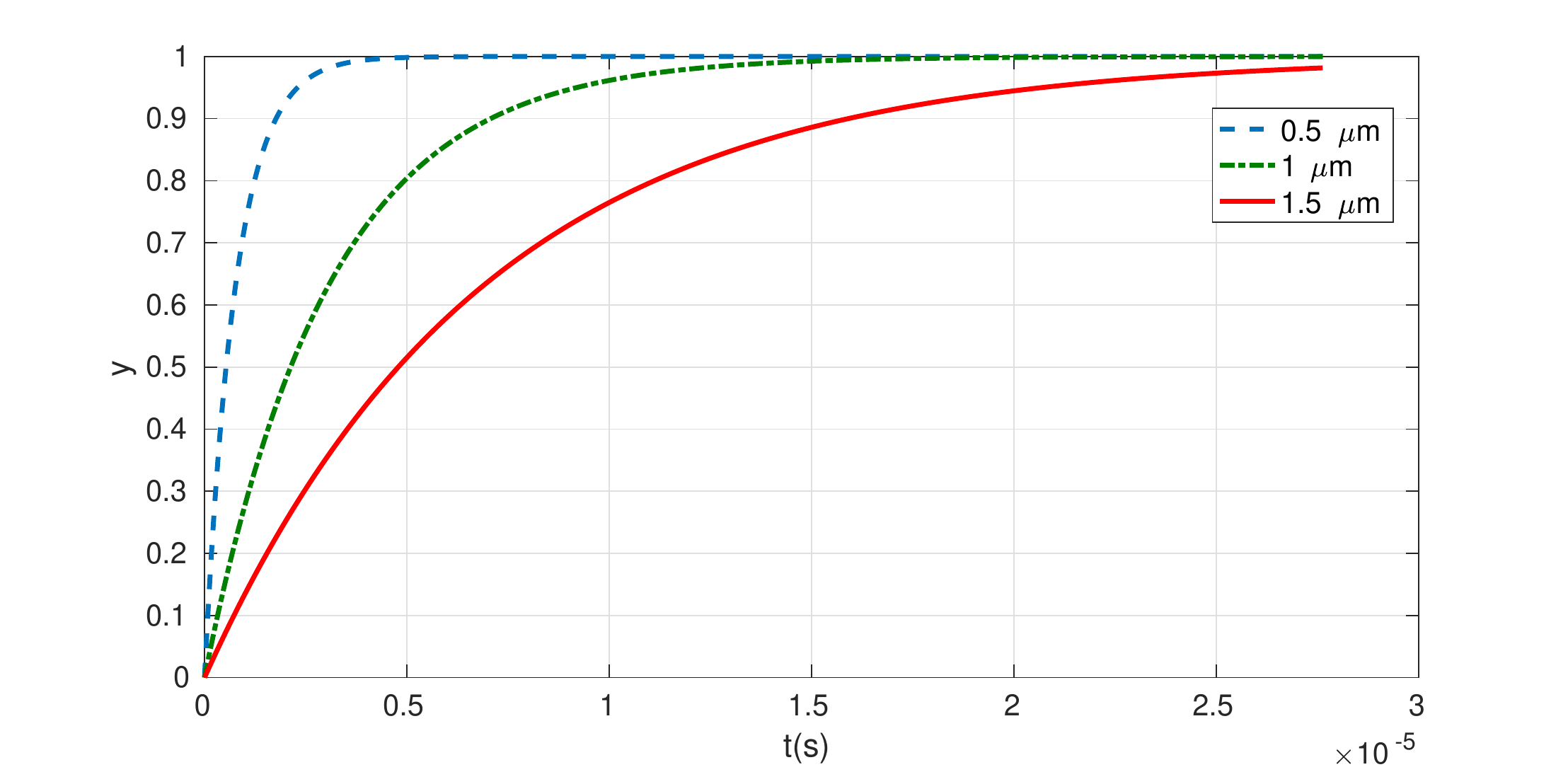}
\caption{The~temporal progress of the fractional velocity, $y=u_p/u_*$ ($u_*=10$~m/s), gain for aerosol particles of different dimensions dragged by the protecting air flow at Stocks regime.}
\label{fig:stocks}
\end{figure}

The~smallest aerosols, up to size $D_L=1.5$~${\upmu}$m, should move in the Stocks
regime with a fractional velocity gaining rate
given by Equation (\ref{sol1}). For the above benchmark parameters of the barrier, the temporal progress of the fractional velocity gain
is displayed in Figure~\ref{fig:stocks}. One can see that the biggest Stocks driven aerosol particles will
reach 90\% of the barrier's flow velocity, $u_p=9$~m/s,
within 16~${\upmu}$s (the corresponding rate is indicated by the solid line
plotted for particle size $1.5$~${\upmu}$m). Let us adjust the inclination angle
in a way that the outward horizontal velocity gaining component would compensate for the inward running velocity,
so that $u_w\approx u_{ph}$. This implies that $\sin\alpha\approx u_w/u_{p}$, so that $\alpha\approx 9.6^{\circ}$.
We can naturally assume that such compensation is maintained with accuracy up to the background speed
of the indoor air, $\Delta u_{ph} = u_w-u_{ph}\lesssim 0.2$~m/s. Thus, the particle being immersed into the air barrier
will be able to leave its width in about $\Delta t_{w}\approx w/\Delta u_{ph}\approx 5$~ms, which can be mapped
into the variation of the inclination angle, $\Delta\alpha\approx 1.15^{\circ}$. Herewith, the vertical
component of the particle velocity will reach $u_{pv}=8.9$~m/s and hence the particle
would be displaced $d_{\rm dsp}\approx u_{pv}\Delta t_w\approx 44$~cm downward. Thus, we expect that the particle will
sweep out downward, to avoid a potential contact with the face openings of the host. If the host is moving with
lower velocity, $\Delta u_{ph}$ can be negative, so that an inward running particle will even be reflected
in addition to the downward displacement described above. Similar reflection might take place in the case
of bigger inclination angle; however, in practice, it is better to maintain it at an optimal value
$\alpha\approx 10^{\circ}$ to minimize spurious horizontal contributions of the barrier flow into the
environmental background air streaming.

\begin{figure}
\centering
\includegraphics[scale=0.7]{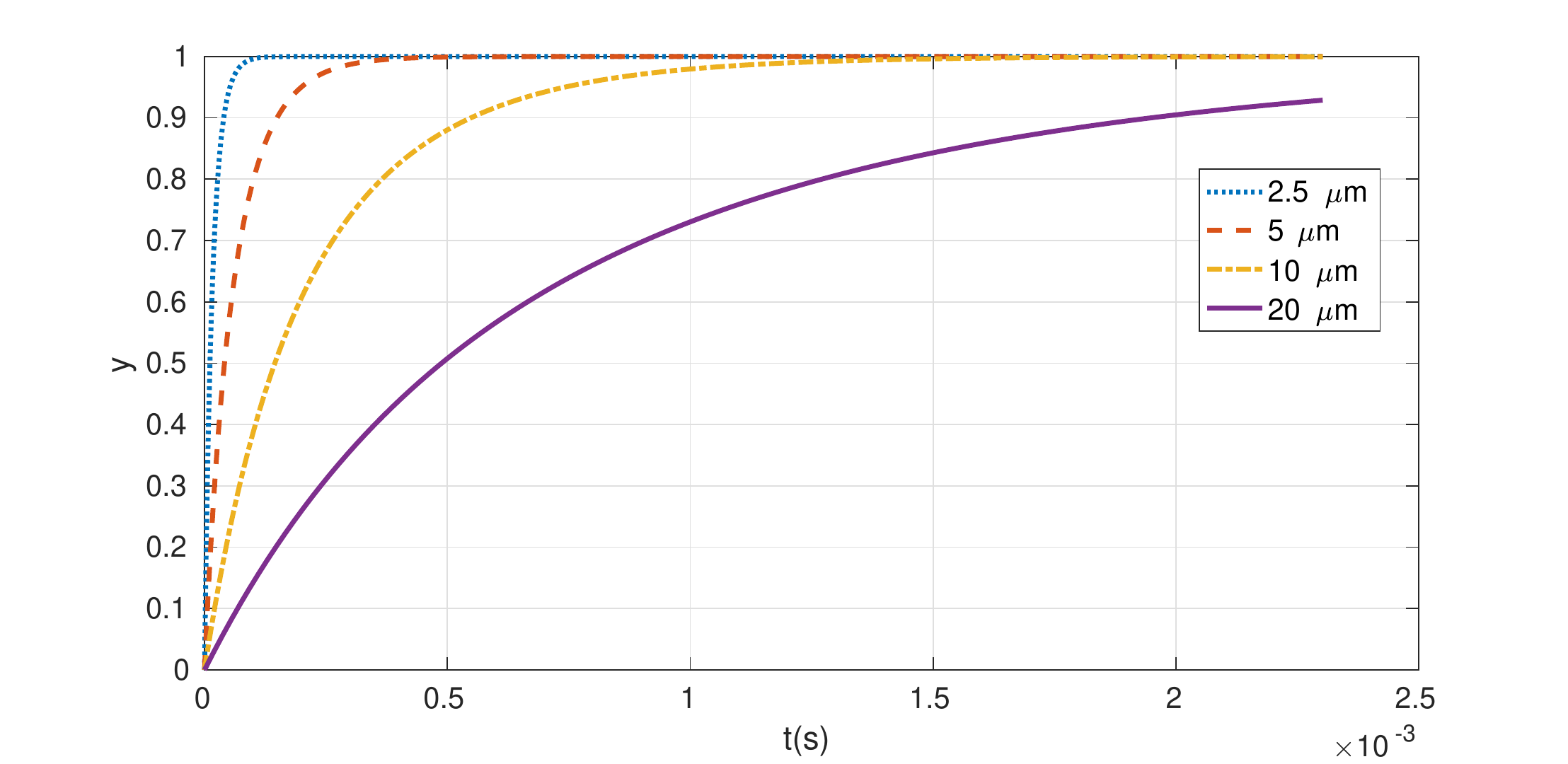}
\caption{The~same as in~Figure~\ref{fig:stocks} plotted for aerosol particles moving at intermediate regime.
Aerosols~of these sizes are still airborne transmittable over long distance.}
\label{fig:interM}
\end{figure}

Bigger aerosols, up to $20$~${\upmu}$m, which might still be suspended in the indoor environment for a long time~\cite{modAir},
will be moved in an intermediate regime in Equation (\ref{sol4}) at slower time progress of the velocity gain, as shown in
Figure~\ref{fig:interM}. Again, as one can see, the airborne aerosols in an intermediate barrier dragging regime
will be accelerated up to 90\% of the barrier's flow velocity within less than in 2~ms.
The~corresponding rate is indicated by the solid line plotted for particle size $20$~${\upmu}$m.
Since the velocity gain is still fast enough, i.e., the full acceleration will be achieved well before the
elapsing of $t_w$, all arguments given in the above paragraph  are valid for the biggest still airborne transmittable particles.
These particles will be successfully displaced ($\gtrsim$44~cm) from trajectories potentially crossing the
face.

The analysis above testifies that the air barrier is able to effectively mitigate inward propagation of
airborne transmitted virus-laden fluid particles of all possible sizes.

We note that smaller particles, $\le$5~${\upmu}$m, might carry higher concentration of the virus than larger
ones~\cite{smallVirus_1,mask_1} and moreover can be transported deep into the lungs, avoiding defence of the
upper respiratory system~\cite{smallVirus_2,inflCovid1}. Also, small aerosols inoculation has been shown
to cause more severe symptoms than bigger particles administered by intranasal inoculation and the
dose of influenza required for inoculation by small aerosols route is 2--3 orders of magnitude lower than that
required by intranasal inoculation~\cite{smallVirus_1}.

\begin{figure}
\centering
\includegraphics[scale=0.7]{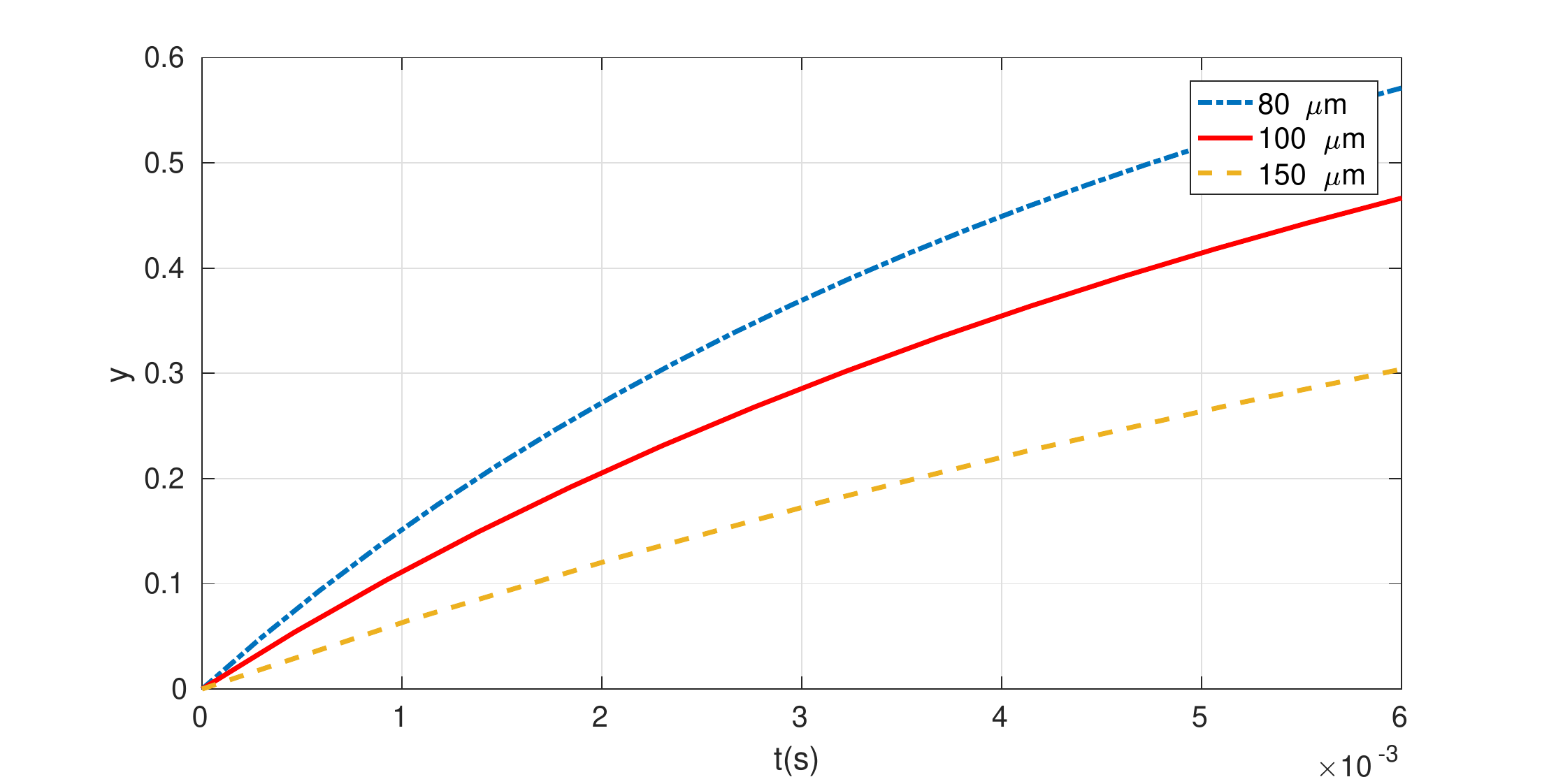}
\caption{The~temporal progress of the fractional velocity gain for droplets whose
sedimentation time exceeds their evaporation time.
The~evaporation indicates the transformation of droplets into airborne aerosols.
The~drug force for such droplets is driven by an intermediate regime.}
\label{fig:interM10}
\end{figure}

The~protection is effectivity degraded for droplets which still can be transformed into airborne
aerosols through evaporation, namely whose sedimentation time starts to exceed their evaporation time.
Thus the smallest size droplets $80$~${\upmu}$m which will still be able to reach the ground
before evaporation~\cite{modAir} gain half of the velocity of the barrier stream within $t_w$.
This is indicated in Figure~\ref{fig:interM10}, by the dashed dotted line, i.e., a particle of such dimension
starts to move with velocity $u_p\approx 5$~m/s after 5~ms. Therefore, the expected
downward displacement will amount to $\dsp\approx 22$~cm. 100~${\upmu}$m and 150~${\upmu}$m
droplets will be displaced by $\dsp\approx 17$~cm and  $\dsp\approx 11$~cm respectively.
Although, the displacements are not enough, except, probably, for $80$~${\upmu}$m droplets,
such particles, being displaced by the air barrier, proceed to fall freely in still air following their
equation of motion, which is defined by gravity $g=9.81$~${\rm m/s^2}$ and kinematic time scale
given in Equation (\ref{rel1}). For example, a $80$~${\upmu}$m droplet should reach its
terminal velocity $u_g=t_{S*}g\approx 0.2$~m/s within $t_{S*}\approx 2$~ms and hence
would meet the ground after $t_g=h/u_g\simeq 8$~s from the moment it was expelled at a hight of
$h=1.6$~m. A bigger droplet, of $150$~${\upmu}$m size, will terminate its acceleration at
$u_g\approx 0.8$~m/s, which implies its sedimentation time $t_g\simeq 2$~s. Therefore,
we believe that it is very likely that droplets of such sizes being moved down by the
air barrier will proceed to settle to the ground instead of coming in a contact
with the host's face. However, one should say, that the definitive conclusion on the effectivity of this kind
of stimulating sedimentation can be established in numerical simulations.

Outward protection is also an important issue in the COVID-19 pandemic, since SARS-CoV-2
transmission may occur earlier in the course of infection, from asymptomatic and
minimal symptomatic hosts, which may expel fluid particles in the range from aerosol size up to a few millimeters.
In the case of the air curtain, the outward protection, for outgoing droplets, is provided due to a kind of
stimulation of ground settling of outgoing particles described above. Moreover, bigger outgoing droplets, of about
$\gtrsim$1~mm size, will be moved by the barrier gas-particle flow with respect to the auto-model
regime in Equation (\ref{sol5}), with the kinematic time scale given by Equation (\ref{rel3}). This implies that,
within $t_{T*}\approx 0.36$~${\upmu}$s (for $D=1$~mm), the droplet will reach half of the
speed of the barrier, namely $u_p=5$~m/s, to be pointed almost vertically down. Thus
it will be ground settled directly around infected individual within 0.3~s instead of
following its ballistic trajectory which can spread 1-2 meters along horizontal direction
in violent expiratory events like coughing and sneezing~\cite{jet_1,jet_2,jet_3}.
Since outgoing aerosols are also displaced by the air barrier, up to 0.5~m downwards, we believe
that it will  suppress the aerosols transport over long distances with indoor air flow.
However, again, for definite answer, numerical simulations, like for example~\cite{modAir},
would be required.

\section{Concluding remarks}
\label{concl}

In the current respiratory COVID-19 pandemic, airborne virus-laden aerosols are considered as the primary route of transmission of
of SARS-CoV-2 pathogen. Staying in the air for a long enough time (minutes or hours), aerosol
particles can be transferred over long distance in an indoor environment~\cite{dropletDry_1,dopletDry} to be
inhaled by a susceptible host.
Face cover masks are currently treated as the best accepted
PPE in mitigating aerosol dispersal. Thus, iregarding mitigation of the virus transmission,
face masks have become a norm in our everyday lives. However, wearing face masks may become uncomfortable
in some situations like in summer heat, while staying on beaches or at hotel swimming pools, doing exercises in gyms, etc.
Moreover, the material barrier of masks or respirators makes it harder to take in air, which might be not be acceptable
for an individual with a chronic respiratory condition like asthma, COPD, etc.

Therefore, we performed the study of a PPE based on non-material protective barrier created by a
flow of well directed down stream of air across the front of an open face.
Unlike in the case of material based protection, such as face masks and respirators, which trap virus-laden
fluid particles via the combined effects of diffusion, inertial impaction, interception,
and electrostatic attraction~\cite{filtr_1}, the air curtain PPE dynamically displaces
the trajectories of infected particles that would otherwise be inhaled by an uninfected
individual from reaching the surface and the openings of the host's protected face. In our analytical study
of the benchmark setup we demonstrated that the inward protection effectiveness of the air barrier
is very high for airborne aerosols of all possible sizes.

For the sake of fairness and consistency, one should say that, unlike in the case of material-based
face covers, the air barrier does not provide the trapping of outward going fluid particles. However, instead,
it stimulates the ground settling of big droplets very close to the infected host and
displaces exhaled aerosols up to 0.5~m downwards and hence, very likely, suppresses the airborne transport
of the infection over long distances with indoor air flow. Moreover, while a mask's material can significantly
reduce the velocity of the through flow jet during expiratory events, presence of the same material leads
to an increase of pressure in the region between the mask and the face and hence results in an increased
perimeter leakage in the form of side jets, as shown analytically and numerically in~\cite{maskAnalysis_1}
and investigated experimentally in~\cite{mask_2}. The~impact
of this process depend on fluid structure, the structural design of the mask
as well as the permeability of the mask's material. The~leakage jets
that are ejected from the perimeter can be turbulent and highly directed (see, for example,
the flow visualizations presented in~\cite{mask_2}), potentially serving as effective dispersers
of respiratory aerosols in transverse directions. Spasmodic expiratory events such as
coughing and sneezing that generate high transient expulsion velocities will significantly
diminish the outward protection effectiveness of face masks~\cite{mask_2}. The~air barrier
does not suffer from the above disadvantages just because there is no physical cover material
which otherwise would cause overpressure and perimeter leakage. Instead, the air barrier's
stream can substantially spoil the structures of jets generated by spasmodic expiratory events.
The~details of the outgoing fluid structure interaction with the air barrier is a matter
of numerical modeling, in a manner of~\cite{maskAnalysis_1}, and as well as, perhaps, experimental studies developed in~\cite{mask_2} and other
dedicated publications.

Our theoretical study of the benchmark setup is aimed to focus the attention of relevant community on the
principal ability of a potentially
portable air curtain to serve as an effective PPE in mitigation of the human to human transmission of
respiratory virus infection like COVID-19. For~an equipment realization and instrumentation we rely
on the current development of technology for construction of portable, low noisy and
low consuming air steaming engines. However, some general attractive every day life advantages of the air curtain PPE
may be elucidated already now. Indeed, there is no necessity of hand washing and disinfection
like in case of putting on and taking off the face masks. Instead, it is supposed that the
air streaming equipment can be simply powered on and off.
This also means that the face washing procedure, which is also important for the removal of
sedimented pathogens in the vicinity of mucus and conjunctiva, is much more simplified
compared to the wearing of face masks, respirators or face shields.
It is difficult to manage to cover the mouth and nose with a flexed elbow or tissue,
when coughing or sneezing with a face mask or respirator on, in particular if a face shield as a PPE
is installed as well. It is clear that this is not the case for an air curtain PPE-wearing individual.
The~interior of the inner instrumentation of the equipment can be disinfected by installing
a miniature interior UV source to be activated while the protecting device is taken off,
so that the unit is self-disinfected.
Although the stream of the barrier is formed out of surrounding air which itself can
contain virus-laden fluid particles, their trajectories will always lie along the plain
of the barrier and hence will never come in a contact with mucus or conjunctiva of
the protected face. Thus, there is no need to instrument the device with a kind of on-flight
disinfection equipment.

\end{document}